# Enhanced UV Photodetector Efficiency with a ZnO/Ga$_2$O$_3$ Heterojunction

Shashi Pandey[1], Swaroop Ganguly[1], Alok Shukla[2*], Anurag Tripathi[3]

[1]Department of Electrical Engineering, Indian Institute of Technology Bombay, Powai, Mumbai 400076, India

[2]Department of Physics, Indian Institute of Technology Bombay, Powai, Mumbai 400076, India
*Email:* shukla@phy.iitb.ac.in

[3]Department of Electrical Engineering, IET Lucknow, Uttar Pradesh 226021, India

**Abstract**

Heterostructures comprising uncoated ZnO and coated with thin layers of Ga$_2$O$_3$ were produced using spin-coating and subsequent hydrothermal processing. X-ray diffraction examination verifies the structural integrity of the synthesized heterostructures (HTs). Optical and photoluminescence spectra were recorded to assess the variation in absorption and emission of the Ga$_2$O$_3$-coated HTs in comparison to the pristine ZnO. We conducted comparative density-functional theory (DFT) computations to corroborate the measured band gaps of both categories of HTs. To assess the stability of our devices, the transient response to on/off light switching under zero bias has been studied. The rise time $\tau_{r1}$ ($\tau_{r2}$) is 2300 (500) ms and the decay time $\tau_{d1}$ ($\tau_{d2}$) is 2700 (5000) ms have been observed for bare ZnO and ZnO/Ga$_2$O$_3$ HTs, respectively. A significant amount of change was also observed in the electrical transport properties from bare ZnO to ZnO/Ga$_2$O$_3$. To see the performance of device, responsivity (R) and detectivity (D = 1/NEP$_B$) have been measured. It is evident from observation that responsivity of a device shows maximum value in UV region while it is reducing with visible region for HTs. In case of detectivity, the maximum value reached was $145 \times 10^{14}$ Hz$^{1/2}$/W (at ~ 200 nm) and $38 \times 10^{14}$ Hz$^{1/2}$/W (at 300 nm) for Ga$_2$O$_3$ coated ZnO, and bare ZnO HTs, respectively. The maximum responsivity measured for the bare ZnO HTs is 7 (A/W) while that of Ga$_2$O$_3$ coated ZnO HTs is 38 (A/W). It suggests a simple way of designing materials for fabricating broad-range cost-effective photodetectors.

**Introduction**

Wide bandgap semiconductors burst into prominence with the application of nitride semiconductors for blue and white LEDs. In recent times, wide bandgap oxide semiconductors have engendered tremendous interest, due to their optoelectronic properties, photocatalytic activity, energy storage capacity, low-cost non-volatile memory functionality, and so on [1-4]. Optoelectronic devices such as LEDs, laser diodes, photodiodes and solar cells involve the interaction of light (photons) and matter (electrons and holes) [5-10]. Among the structures that can most benefit from material nano structuring to enhance their response to light stimuli are photodetectors [11-15]. Although nanostructures have enormous potential, their primary drawback is that they require an external bias voltage to function [15]. This limits the range of applications that can be made of them, including in situ medical therapy monitoring, nanorobotics, wireless environmental sensing, security applications, and portable personal electronics [16–19]. Ultra-wide bandgap (UWBG) oxide semiconductors have been considered for use in next-generation electronic devices due to their potential uses in bioelectronics, solar-blind detection, and strategic applications in recent years [15,20–23]. Utilizing the photovoltaic (PV) effect in the device structure which produces electron-hole pairs in semiconductors when exposed to light self-powered photodetectors can be created [1,24,25].

However, it is important to recognize that the primary issue with UWB materials is the inability to accomplish concurrent n- and p-type doping inside them. Nonetheless, gallium oxide ($Ga_2O_3$) is a promising UWB material that enables the production of cost-effective, melt-grown substrates on a wide scale with low defect density, in addition to facilitating adjustable n-type doping [26,27]. Owing to its substantial band gap of around 4.9 eV and exceptional physical and chemical stabilities, it is utilized in deep-UV photosensors as well as high-power devices. $Ga_2O_3$-based power electronic devices are expected to offer enhanced blocking voltage with reduced specific on-resistance for power switching applications [11,28–30]. Substantial progress has been achieved in recent years due to rigorous study focused on investigating its material properties and evaluating the limitations of devices constructed from it. Consequently, $Ga_2O_3$ is regarded as one of the most promising materials for next-generation high-power and high-efficiency devices [30–32].

Self-powered wideband photodetectors have garnered significant interest recently for applications in various fields, including portable electronics, implanted biological sensors, secure communication, and environmental monitoring [33–35]. Metal oxide nanostructures and their structural, optical, and electrical properties are garnering significant attention; yet, fabricating diverse optoelectronic devices with varied metal oxide combinations while achieving optimal performance can be challenging. $Ga_2O_3$, $Zn_2GeO_4$, ZnO, $TiO_2$, and $SnO_2$ are among the several wide bandgap metal oxides that have been explored for application in heterojunction-based UV/visible detectors [30,36–41]. ZnO-based devices are the optimal selection for next-generation photodetectors owing to their superior thermal and chemical stability. In photoconductive mode, such detectors have comparatively shorter response times and diminished responsivity, while adequate detectivity cannot be achieved without additional power. Consequently, self-sustaining, high-efficiency detectors are greatly sought after for future portable, cost-effective applications. Modifying the barrier heights and band offsets of the heterojunctions might adversely affect photo-carrier transport, resulting in significantly low responsivities. Various heterojunctions, such as $Ga_2O_3/SnO_2$ and $Ga_2O_3/ZnO$, have been suggested for the development of self-powered detectors exhibiting significant responsivity [11,40]. Consequently, substantial effort will be necessary to improve the detector's performance by increasing the interfacial area, thereby enhancing light absorption. According to Daotong *et al.*, the fabricated device showed exceptional photoresponsivity characteristics with a large photo-to-dark current ratio of $2.64 \times 10^4$, detectivity (D) of $6.11 \times 10^{12}$ $Hz^{1/2}W^{-1}$, and a responsivity (R) of 137.9 mA $W^{-1}$ [11]. Xie et al. investigated the β-$Ga_2O_3$/ZnO nanocomposite heterogeneous structure using sputtering method for its employment in UV sensing of improved performances [15,16].

This work examines the optical properties of crystalline bare ZnO nanostructures and those coated with thin layers of $Ga_2O_3$ (ZnO/$Ga_2O_3$). We selected these heterostructures due to the inherent deep levels and surface defects in pure ZnO, which exacerbate unwanted photocurrent and result in diminished sensitivity and unreliable switching during device operation. Consequently, pure ZnO is an inadequate choice for such devices [10,17–19]. The conductivity, transparency, and defect-related energy transitions in ZnO can be easily modified by the addition of dopants [11,19–21]. However, below a certain threshold, foreign atoms disrupt the crystalline structure of ZnO due to lattice scattering and may increase the probability of recombination of photogenerated carriers [17,22]. Surface flaws significantly impact device performance, and addressing these defects can greatly enhance detector efficacy through several methods, including surface plasmon coupling, the decorating of metal nanoparticles on nanostructures, and the fabrication of core-shell nanowires. Thus, the core-shell architecture in ZnO nanowires probably reduces surface defect-induced recombination and improves photo response. A variety of semiconducting materials, such as $Cu_2O$, $SnO_2$, ZnSe, and $TiO_2$, have been combined with ZnO to create core-shell nanowires [40,42–44]. The predominant core-shell architectures employing ZnO operate within the UV-visible spectrum, demonstrating poor responsivity, high operating voltage, and restricted detection bandwidth. Thus, we utilized $ZnO/Ga_2O_3$ heterojunctions, distinguished by a wider energy bandgap, increased exciton binding energy, and little lattice misfit, making them more suitable for the development of core-shell heterojunction photodetectors.

**Experimental Method**

**A. $ZnO/Ga_2O_3$ Heterostructure (HTs): Synthesis and Characterization**

First, acetone and methanol or IPA were used to clean the indium-tin oxide (ITO) substrate by initially mixing it in acetone, and then, as soon as it was removed from the acetone, it was quickly put into methanol before it dried. Then, 25 mL of 2-methoxy ethanol was used to dissolve the 1:2 combination of zinc acetate dehydrates and ethanolamine. The prepared solution was then stirred with the help of spin coating at 50 $^0$C for an hour with 2500 rpm for one minute onto ITO glass surfaces. To create ZnO nanostructures on the ITO matrix, the resulting sample was heated for an additional six hours at 350 $^0$C. In a continuous procedure, diluted hydrochloric acid was used to dissolve in commercial $Ga_2O_3$ (99.99%) and make a solution. After evaporation, a clear solution was created by adding 35 ml of deionized water or a mixture of diethylene glycol (DEG) and water (DEG: $H_2O$ 1:1 vol). The agitated solution was then added dropwise with 1 $mol^{-1}$ of sodium hydroxide solution till $P_H$= 1/4. As a result of this, a colloidal precipitate was obtained, and then transferred into a autoclave along with prepared ZnO nanostructures on ITO, and heated at 180 °C for 24 hours. After that, the autoclave was cooled to ambient temperature at a rate of roughly 1 K $min^{-1}$. Following filtration, washing, and drying at 60°C in air for 6 hours, the hydrothermal crystals were obtained. To validate the deposition of crystalline phase $Ga_2O_3$ on ZnO HTs, the washed sample was subjected to annealing at 450 °C for 6 hours. Powder X-ray diffraction (XRD) tests were conducted using a Bruker D8 diffractometer with a Cu target and a LYNXEYE detector to examine the structural phase purity of the generated samples. The structural purity of the produced samples was verified using high-temperature X-ray diffraction studies. Diffuse refraction measurements have been used to determine the prepared sample's optical band gap. These measurements were made in Perkin Elmer UV-Vis-NIR Spectrophotometer with 200 nm to 800 nm wavelength region. Using thermal evaporation, an electrode made of Au (thickness 50 nm) has been deposited for the preparation of devices. Masking was employed with the width limit of 1 mm and the channel length of 100 mm (distance between Au electrodes). The thickness of bare ZnO was around 100 nm and for HTs it was close to 150 nm. The Keithley 2612A source meter was used to perform the electrical characterizations [10,23-25].

## B. Computational details

First-principles calculations were carried out within the framework of plane-wave density-functional theory (PW-DFT) as implemented in the Vienna ab-initio simulation package (VASP) [41,45-47] to support the experimental findings. The generalized gradient approximation and the Hubbard U level of theory (GGA+U) [7,8,48,49] were used to validate the calculated band gap with the experimentally measured one. We used a supercell of bulk ZnO with the dimensions 3x3x1 and added 4 Ga and 8 O atoms to one of the edges of the ZnO super cell to model the ZnO/$Ga_2O_3$ HTs (see Fig. 1) Next, the geometries of both the pure ZnO and $Ga_2O_3$-coated ZnO supercells were optimized, and iterations were terminated only when the force experienced by each atom declines to less than 0.04 eV/A°, and the system's total energy is stable within $10^{-3}$ mRy. Regarding self-consistency, when the energy convergence reaches up to 5x$10^{-5}$ eV, the self-consistent computations are considered to be converged [3,6,8,50,51]. Figure 1 (c) shows the schematic view of fabricated HTs of ZnO/$Ga_2O_3$ (discussed in synthesis section).

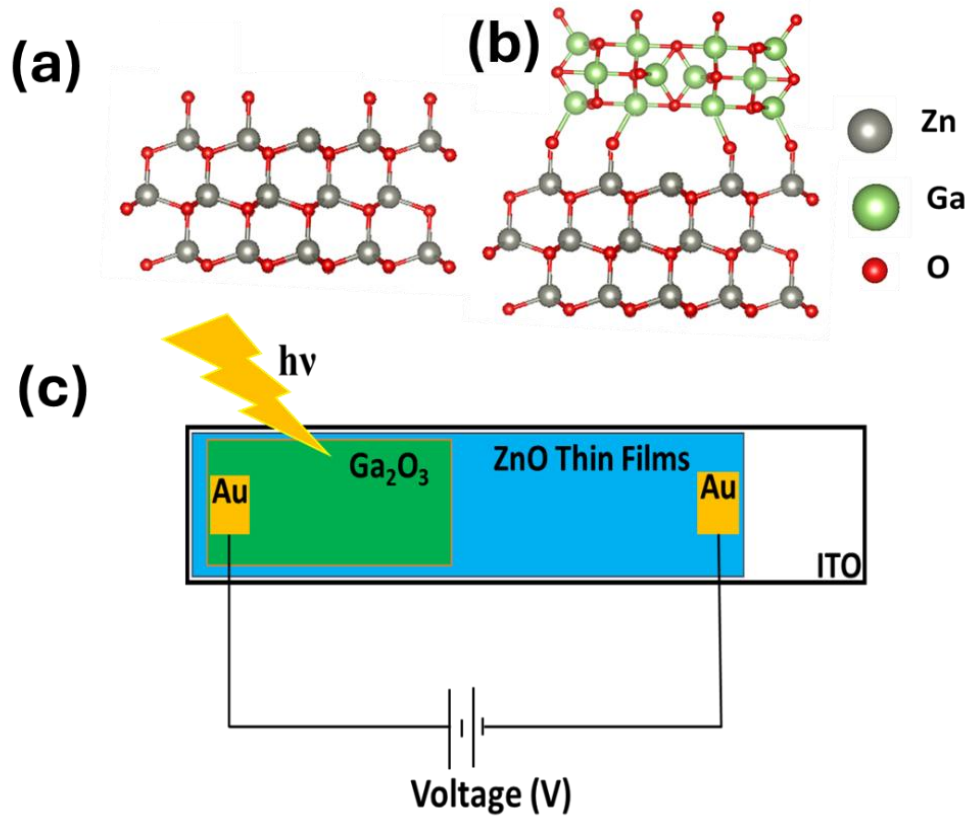

**Figure 1:** S*tructures of (a) bare ZnO and (b) ZnO/Ga₂O₃ nanocomposites (c) Schematic of fabricated HTs of ZnO/ Ga₂O₃.*

Results and Discussion

We performed X-ray diffraction to determine the crystal structure and stability of our deposited bare ZnO and ZnO/Ga$_2$O$_3$ nanocomposites (see Figs. 2(a) and (b)). Fig. 2(a) displays the X-ray diffraction peaks of its corresponding hkl planes of as-prepared ZnO nanostructures. All diffraction peaks supports with indexing of hexagonal wurtzite structure of ZnO [1]. Figure 2(b) depicts the diffraction pattern of ZnO/Ga$_2$O$_3$, and additional peaks of monoclinic phase of Ga$_2$O$_3$ are clearly visible (indicated in blue color). Ga$_2$O$_3$ has additional peaks corresponding to (200), (400), and (600) planes that are consistent with its monoclinic crystal structure. The creation of the ZnO/Ga$_2$O$_3$ composite HTs is indicated by the coexistence of the ZnO and Ga$_2$O$_3$ phases. Figs. 2(a) and 2(b) show the absence of any additional peaks and which confirms the deposition of only Ga$_2$O$_3$ on ZnO nanostructures.

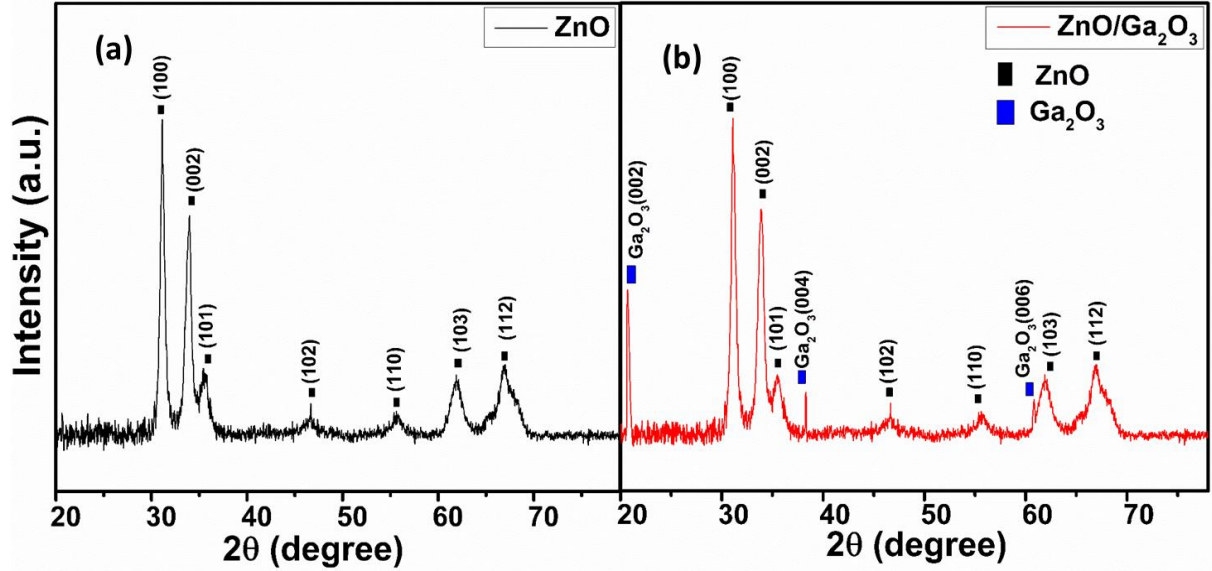

**Figure 2:** *X-ray diffraction patterns of: (a) bare ZnO, and (b) ZnO/$Ga_2O_3$ HTSs. Planes (002), (004) and (006) of $Ga_2O_3$ are shown in the blue color in (b).*

We employed diffuse reflectance spectroscopy to assess the optical absorption spectra of the examined nanostructures, aiming to determine the optical gaps and absorption profiles of deposited bare ZnO and ZnO/$Ga_2O_3$ heterostructures across the wavelength range of 200 nm to 800 nm [35–39]. The optimized geometries of the Bulk ZnO nanostructures were found to be 39.28 Å long, with a diameter of 5.20 Å [1]. Bulk ZnO is recognized to possess a band gap of 3.30 eV, whereas our measured optical gap for bare ZnO nanostructures is approximately 3.58 eV (see Fig. 3(a)), which, in all likelihood, is due to quantum confinement effect in transition from the bulk to the nanostructure phase. While $Ga_2O_3$ has a significantly wider band gap as compared to ZnO, i.e., 4.5 eV, however, the optical gap of $Ga_2O_3$-coated ZnO HTs is 4.10 eV (see Fig. 3(a)). Clearly, this difference is due to the fact that a relatively small amount of $Ga_2O_3$ coating has been employed in making the heterostructure. To check the structural stability of bare and HTs, first the formation energy of bare ZnO and was computed using the formula

$$E_{form}(ZnO) = E_{tot}(ZnO) - [\sum_{x=1}^{n} x\, E_{tot}(Zn) + \sum_{y=1}^{m} y\, E_{tot}(O)], \qquad (1)$$

Similarly, for heterostructures:

$$E_{form}(ZnO/Ga_2O_3) = E_{tot}(ZnO/Ga_2O_3) - [\sum_{x=1}^{n} x\, E_{tot}(Zn) + \sum_{y=1}^{m} y\, E_{tot}(O) + \sum_{z=1}^{r} z\, E_{tot}(Ga)], \qquad (2)$$

Above, $E_{form}(ZnO)$ and $E_{form}(ZnO/Ga_2O_3)$ represent formation energies of bare ZnO and ZnO/Ga$_2$O$_3$, respectively, while $E_{tot}(ZnO)$ and $E_{tot}(ZnO/Ga_2O_3)$ denote the total energies of the bare ZnO and ZnO/Ga$_2$O$_3$ unit cells, respectively. Furthermore, $E_{tot}(Zn), E_{tot}(O)$, and $E_{tot}(Ga)$ represents total energies of the individual atoms Zn, O and Ga, respectively, while the atoms are indexed by by x, y and z respectively. Using eqs. (1) and (2), the formation energies have been calculated to be -2.67 and -0.82 eV per unit cell for the bare ZnO and ZnO/Ga$_2$O$_3$, respectively. Note that the negative values of the formation energies indicate structural and thermodynamic stability of a system.

In a ZnO/Ga$_2$O$_3$ heterojunction, the bandgap reduction is mainly due to the alignment of energy bands, interface states, and strain induced by lattice mismatch. The interaction of these two materials results in a modified electronic structure, which can lead to a smaller effective bandgap than in the individual bulk materials.

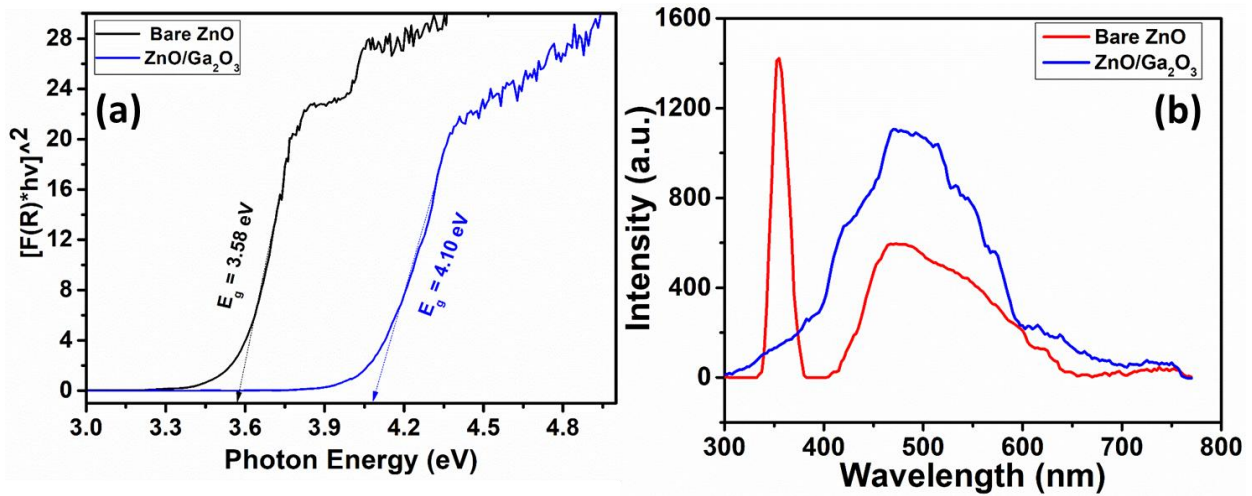

**Figure 3:** *Measured **(a)** Optical absorption spectra and **(b)** Photoluminescence of bare ZnO and ZnO/Ga$_2$O$_3$ HTs.*

We also recorded the PL spectra (see Fig. 3(b)) of the two types of HTs to see to understand the influence of the Ga$_2$O$_3$ coating on the optical properties of the ZnO HTs. From Fig. 3(b) it is clear that in bare ZnO luminescence corresponding to the energy of its band edge (350 nm) and defect states (450 nm-600 nm), while for the ZnO/Ga$_2$O$_3$ HTs, the peak at 350 nm is missing and leads to a broad luminescence spectrum from range 300-700 nm.

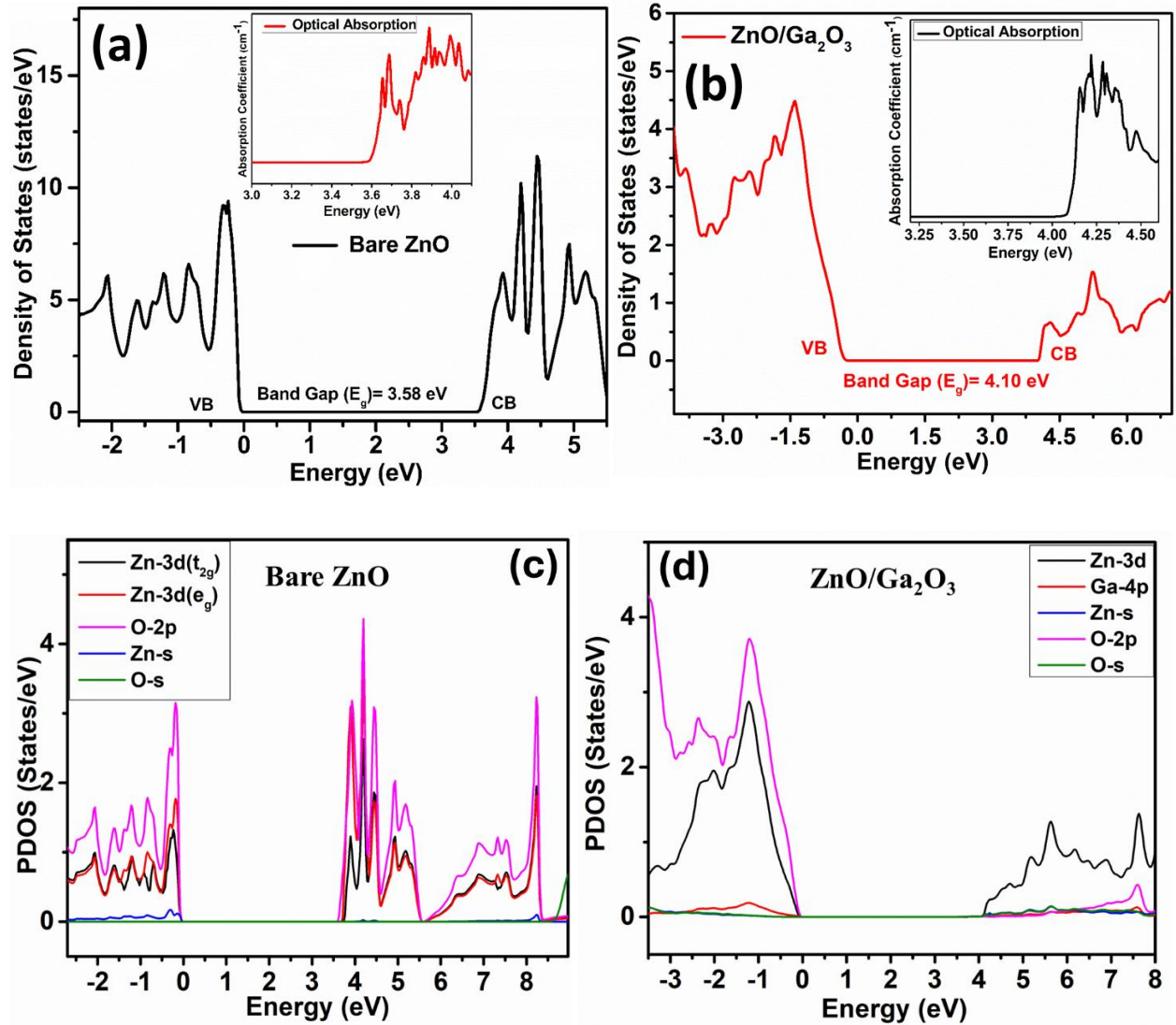

**Figure 4:** *TDOS of (a) bare ZnO & (b) ZnO/Ga$_2$O$_3$ HTs show bandgap of 3.58 eV and 4.10 eV respectively, obtained using the GGA+U (U=5.0 eV) approximations. Inset of Figure 4 (a) and (b) shows simulated optical absorption spectra. Partial density of states (PDOS) calculations shown in (c) for bare ZnO and (d) ZnO/Ga$_2$O$_3$.*

To confirm the experimentally observed bandgap of nanocomposites of ZnO and Ga$_2$O$_3$, we have performed systematic first-principle DFT calculations. In Fig. 4 (a) and (b) we display the calculated total density of states (TDOS) of the bare ZnO and ZnO/Ga$_2$O$_3$, respectively, while the corresponding insets display the calculated optical absorption spectra. For these calculations GGA+U approach was employed, with U=5.0 eV for the Zn-3d orbitals and U=5.8 eV at the Ga-4p orbitals so as to obtain a good match with the experimental bandgaps. The fact that there is a strong agreement between theoretically calculated band gaps and experimentally measured values indicates that we have made the right choices for the values of the Hubbard parameter (U) in the GGA +U (U=5.0 eV: Zn-3d,& U=5.8 eV:Ga-4p) calculations. We also performed partial density of states (PDOS) calculations to understand the contributions of various orbitals to DOS, and the results are presented in Figs 4(c) and 4(d) for bare ZnO and ZnO/Ga$_2$O$_3$. It is clear from these figures that the valance band maxima (VBM) and conduction band minima (CBM) in both cases are due to the hybridization in between O (2p) /Ga (4p) (VBM), and Zn (3d)/Ga (4p) (CBM) orbitals.

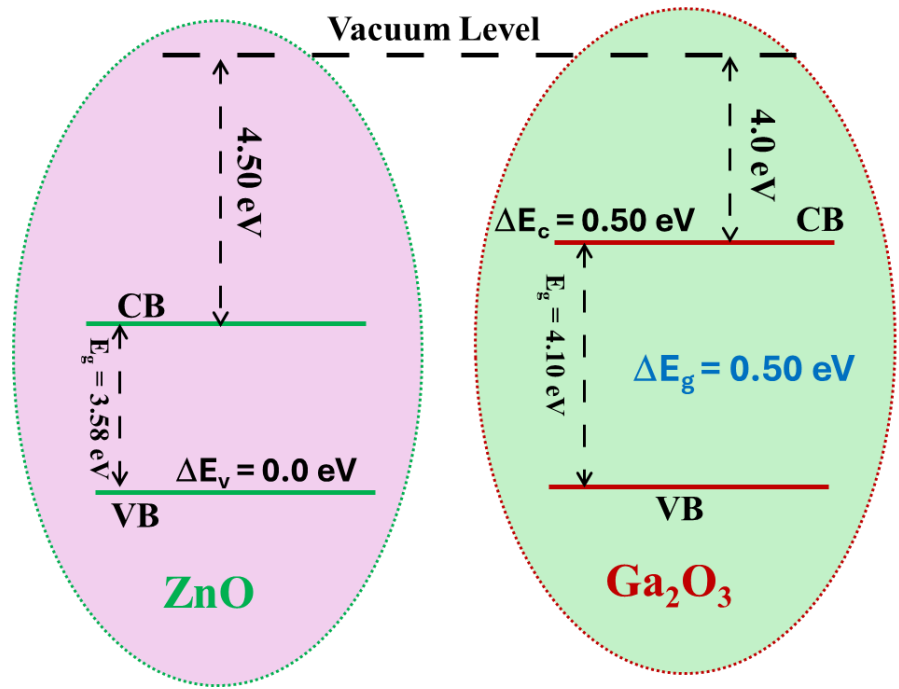

**Figure 5:** *Schematic alignment of band energies in ZnO and Ga$_2$O$_3$ (before junction formation).*

We present band diagrams for the two materials in Fig. 5 to help us understand the working mechanism of the heterojunction photo detector. As shown in the figure, according to our DFT calculations, Ga$_2$O$_3$ and ZnO have electron affinities of 4.00 & 4.50 eV, respectively. The comparable DFT computations, utilizing the derived bandgaps, forecast the conduction band offset ($E_c$) as 0.5 eV and the valence band offset ($E_v$) as 0.12 eV, respectively. Electron-hole pairs are produced through photon absorption when this heterojunction photodetector is exposed to UV light. Consequently, the holes migrate to the valence band edge while the electrons advance to the conduction band edge. The heterojunction photodetector exhibited inadequate rectifying ability due to the formation of a positive peak barrier at the heterogeneous interface. The ZnO/Ga$_2$O$_3$ arrangement presents an interfacial barrier that enables self-powered operation. The valence and conduction band of Ga$_2$O$_3$ would concurrently migrate downward to inhibit excessive hole migration to ZnO. As a result, electrons from the conduction band of ZnO would migrate to Ga$_2$O$_3$, leading to the creation of a junction. Correspondingly, the conduction band of ZnO will ascend to mitigate the excessive efflux of electrons, while the valence band concurrently ascends.

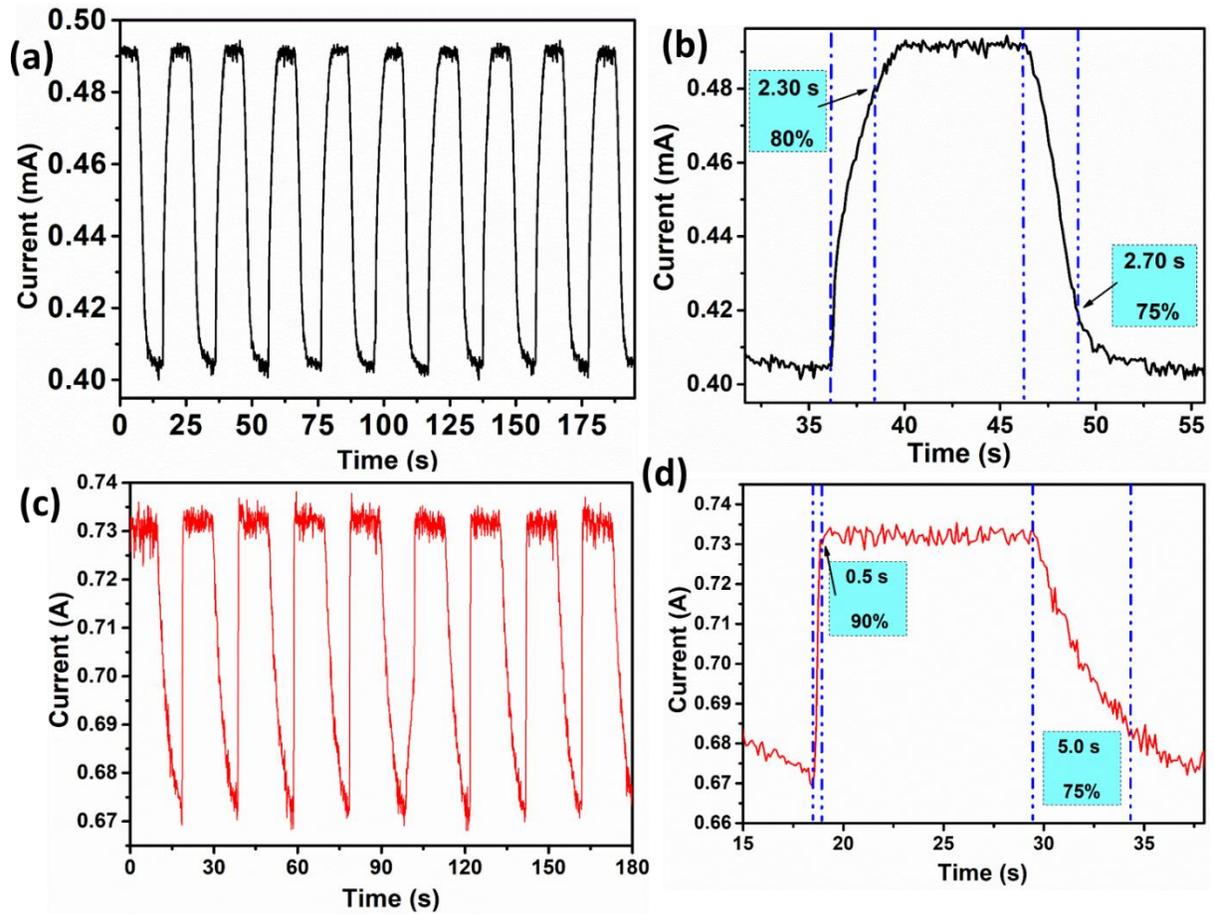

**Figure 6:** *(a) and (b) shows on/off switching response with ultraviolet range for bare ZnO, while (c) and (d) shows response times of ZnO/Ga$_2$O$_3$ heterojunction.*

Our fabricated bare ZnO and ZnO/Ga$_2$O$_3$ heterojunction photodetector works in a self-powered way is undoubtedly verified by the on/off switching under UV range with zero bias, which allows us to determine the stability of our devices and the estimated response time as shown in

Figs. 6(a)-(d) The rise time $\tau_{r1}$ ($\tau_{r2}$) is 2300 (500) ms & decay time $\tau_{d1}$ ($\tau_{d2}$) is 2700 (5000) ms for the bare ZnO and ZnO/Ga$_2$O$_3$ HTs, respectively. The experimental values indicate that the device exhibits efficient photogeneration and recombination. The higher decay time observed in ZnO/Ga$_2$O$_3$ HTs with respect to bare ZnO could be attributed due to having broad PL spectra in case of HTs. For a prepared device, the photocurrent response to the UV light increases from mA to A range in the case of HTs. The repeatability and consistency of cycles confirm the reliability of both prepared devices.

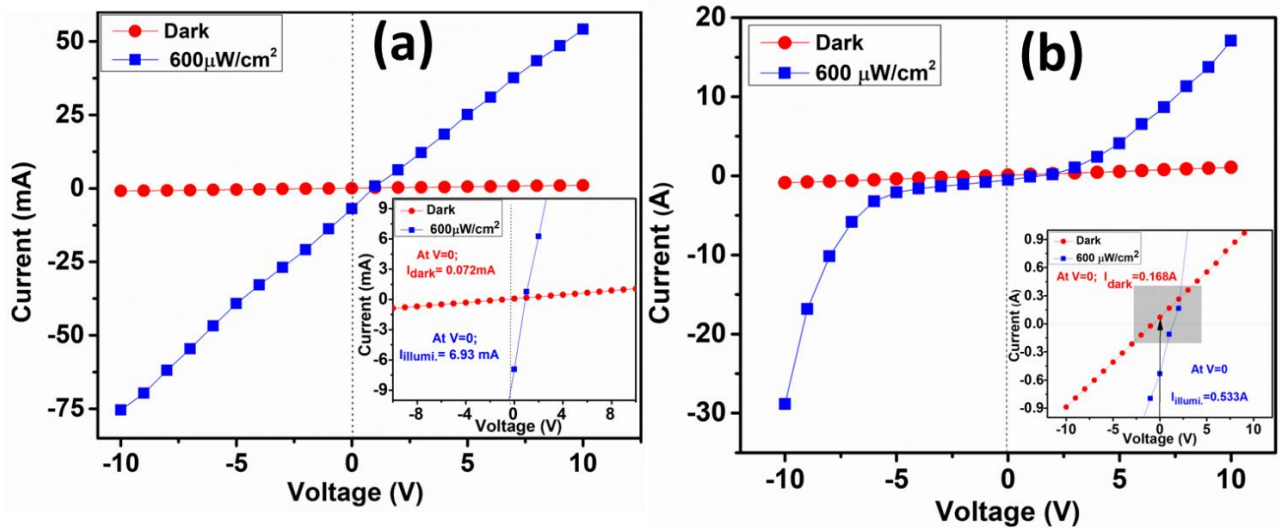

**Figure 7:** *Current-Voltage measurement under dark and illumination with 600 W cm$^{-2}$ using 300 nm wavelength source on: (a)bare ZnO, & (b) ZnO/Ga$_2$O$_3$ heterojunction, while insets of figures (a) & (b) show enlarged views of Current-Voltage measurement at zero bias voltage.*

Figure 7(a) displays the Current-Voltage measurements under dark and illumination with 600 W cm$^{-2}$ using 300 nm photon source. We have a low dark current ($I_{dark}$) of 1.5 nA and a huge photocurrent of 50 mA, for a bias of 10 V in case of bare ZnO nanostructures. However, ZnO/Ga$_2$O$_3$ HTs still show a very low dark current (~5 nA), but under the illumination conditions, the value of current increases many folds (i.e. 16 A) at the same bias voltage of 10V (see Fig. 7 (b)). Hence, the prepared photodetector with gallium oxide heterojunction exhibits excellent device performance. With the 350 nm light illumination intensity of 600 μW cm$^{-2}$, the heterojunction photodetector demonstrates distinct I-V characteristics with an open-circuit voltage ($V_{oc}$) of 1 V for bare ZnO and 2.1 V for ZnO/Ga$_2$O$_3$ HTs. Insets of Fig. 7(a) and (b) respectively show that there are still small amounts of charge carriers inside the device at V= 0, irrespective of whether illumination intensity is there or dark condition. This phenomenon is pointing towards the possibility of obtaining self-powered operation.

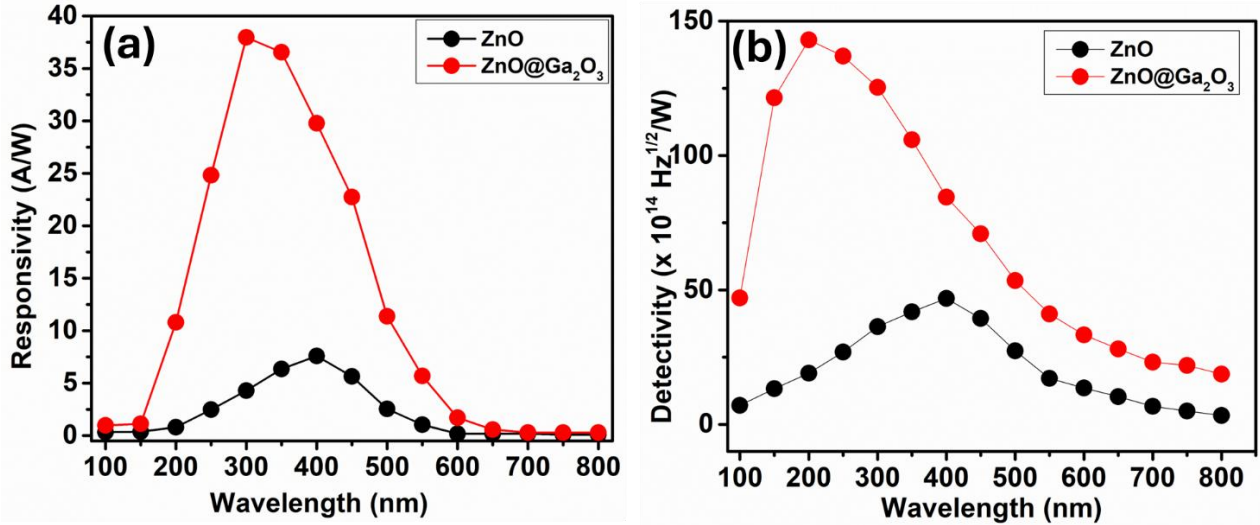

**Figure 8:** (a) *Detectivity, and (b) Responsivity of bare ZnO and ZnO/Ga$_2$O$_3$.*

Responsivity, R = ($I_{UV}/I_P$), is a crucial factor in determining a device's performance. Here, $I_P$ stand for incident power and $I_{UV}$ for maximum current when exposed to the UV-Vis light. From Fig. 8(b) it is obvious that device responsivity is highest in the range of 200-500 nm while reducing for the larger wavelengths for the HTs. Fig. 8(a) shows that the detectivity (D = 1/NEP$_B$) for a photodetector is a figure of merit used to characterize performance, equal to the reciprocal of noise-equivalent power (NEP$_B$) reaches the peak values at $145 \times 10^{14}$ Hz$^{1/2}$/W (at ~ 200 nm) and $38 \times 10^{14}$ Hz$^{1/2}$/W (at 300 nm) for bare ZnO, and Ga$_2$O$_3$ coated ZnO HTs, respectively. Thus, we conclude that the ZnO HTs with Ga$_2$O$_3$ coating far superior device performance with much larger responsivity and detectivity, as compared to the bare ZnO nanostructures.

Whereas, the recombination resistance (R$_{rec}$) is influenced by changes in the recombination flux, the transport resistance (R$_{tr}$) is inversely related to carrier conductivity (σ).

$$R_{tr} = L/A \, \sigma \tag{3}$$

$$R_{rec} = \frac{1}{A}\left(\frac{\partial J_{rec}}{\partial V}\right)^{-1} \tag{4}$$

where A represents the effective cross-sectional area of the device, L is the layer's thickness, V denotes the applied voltage, and $J_{rec}$ is the recombination current.

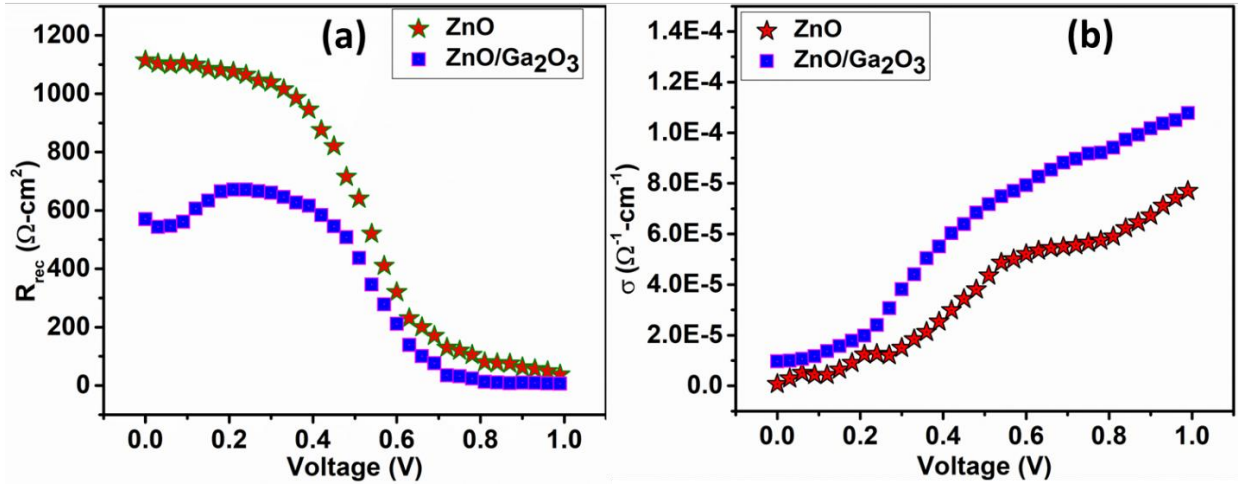

**Figure 9:** *Transport and recombination parameters vs the applied voltage: (a) recombination resistance ($R_{rec}$), and (b) conductivity (σ), of the bare ZnO and ZnO/Ga$_2$O$_3$ heterostructures.*

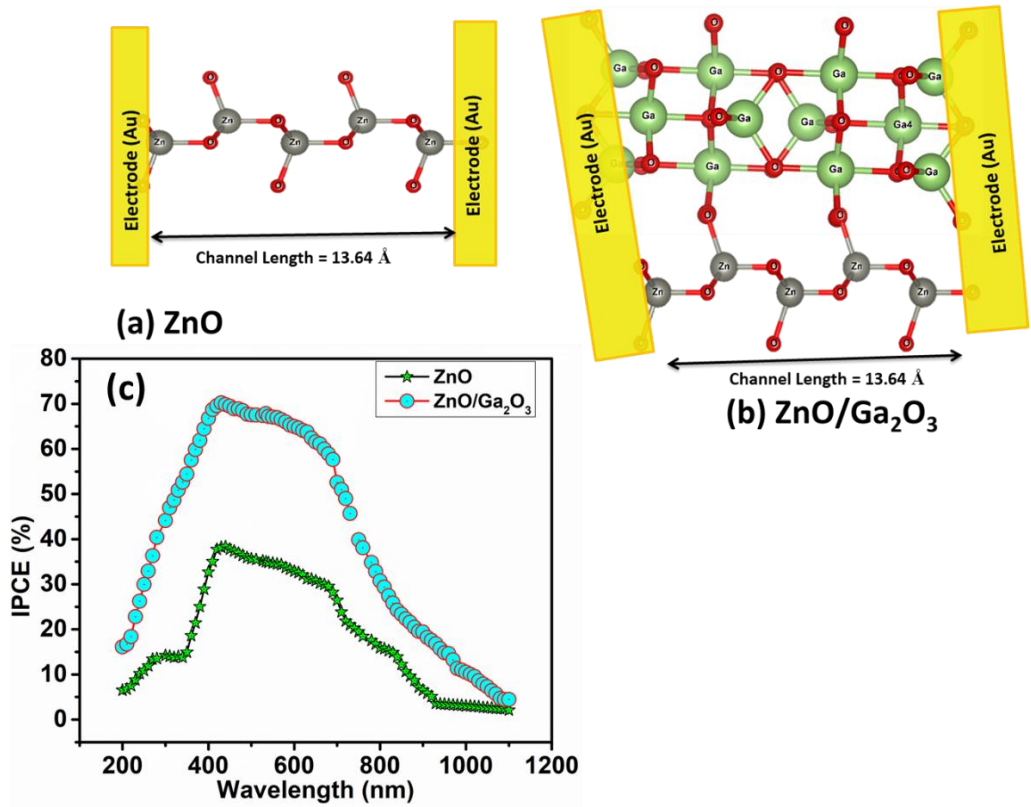

**Figure 10**: *Schematic illustration of the device in different configurations: (a) Pristine ZnO, (b) ZnO/Ga$_2$O$_3$ heterostructure; where red balls indicate oxygen atoms, gray balls represent zinc atoms, green balls represent gallium atoms, and (c) IPCE with respect to wavelengths (150-1200 nm) of the bare ZnO and ZnO/Ga$_2$O$_3$ heterostructures.*

A significant amount of information about the recombination processes in optoelectronics-based devices can be obtained from its recombination resistance, R$_{rec}$, which is shown in Figure 9(a) for our device. For the bare ZnO samples, we can observe that at low voltages, the device has higher recombination resistance, implying higher recombination rates than in the ZnO/Ga$_2$O$_3$ HTs. However, at higher voltages, both have almost the same resistance. Thus, two types of devices exhibit significant changes in recombination rates and, consequently, carrier conductivity as is obvious from Fig. 9(b), which indicates higher carrier conductivity for ZnO/Ga$_2$O$_3$ HTs as comparison to the bare ZnO. The device schematics are shown in Figs. 10a and 10b for IPCE measurement, where Au has been selected as the electrode. In order to quantify the quantum efficiency of the performed ICPE% measurements (see Fig. 10(c)) from which it is obvious that the quantum efficiency shows a notable improvement for ZnO/Ga$_2$O$_3$ HTs device as compared to the bare ZnO device. This is due to the fact that the addition of Ga$_2$O$_3$ on the ZnO sample leads to enhanced white and visible light detection, in agreement with the measurements of Raveesh et al. [12] for the white light detection, and of Jo et al. [13] for the visible light detection.

Additionally, Table 1 summarizes a comparison on performance parameters of Zinc oxide-based photodetectors. Considering this, we observed that ZnO coated with Ga$_2$O$_3$ is one of the most effective self-powered UV photo detectors.

**Table 1:** Performance comparison of our devices as compared to other photodetectors.

| Photo detectors | Dark Current (A) at 0 V | Photo-current (I$_{ph}$) (A) | Response Time (s) | ON/OFF Ratio | References |
|---|---|---|---|---|---|
| ZnO/p-Si | 2.82 x 10$^{-9}$ | 9.35 x 10$^{-3}$ | 111/75 s | 334 | [29] |
| ZnO/TiO$_2$ | 6 x 10$^{-7}$ | 2 x 10$^{-4}$ | 9/20 s | 388 | [1] |
| Y-ZnO/Ga$_2$O$_3$ NWs/p-Si | 1.74 x 10$^{-11}$ | 5.64 x 10$^{-7}$ | 80/88 s | 32.41 | [29] |
| ZnO-Based UV-C Photodetector | 0.19 x 10$^{-6}$ | 3.2 x 10$^{-6}$ | 160/107 s | 2 x 10$^5$ | [9,52] |
| ZnO-ZnCr$_2$O$_4$ nanowalls | 2.6 x 10$^{-6}$ | ~ 10$^{-5}$ | 200/160 | 1.28 x 10$^5$ | [15,53] |
| Bare ZnO | 1.5 x 10$^{-9}$ | 50 x 10$^{-3}$ | < 30 s | 375 | **This Work** |
| ZnO/Ga$_2$O$_3$ | 5 x 10$^{-9}$ | 16 | < 10s | 416 | |

## Conclusion

In conclusion, ZnO samples were produced with thin layers of Ga$_2$O$_3$ heterostructures utilizing the spin coating and hydrothermal methods. X-ray diffraction investigation verifies the structural integrity of the synthesized heterostructures. The responsivity and detectivity of Ga$_2$O$_3$-coated ZnO HTs devices in the UV-Vis spectrum were significantly greater than those of bare ZnO devices, resulting in enhanced photo-detection, sensitivity, and switching stability in the UV-Vis region. A notable disparity has been detected in the electrical transport capabilities of the bare ZnO devices compared to the two devices fabricated from ZnO/ Ga$_2$O$_3$. The performance discrepancies in the fabricated devices (both bare ZnO and ZnO/ Ga$_2$O$_3$ HTs) exhibit substantial alterations in recombination rates and carrier conductivity. Quantum efficiency measurements (IPCE%) indicated a significant enhancement in the results for devices utilizing ZnO/ Ga$_2$O$_3$ HTs in contrast to those employing bare ZnO. Consequently, the use of Ga$_2$O$_3$ on ZnO enhances the device's capability for improved detection of white and visible light. Additionally, first-principles DFT computations were conducted on ZnO nanostructures both with and without Ga$_2$O$_3$ covering, and the results corroborated our experimental observations. Consequently, we assert that heterostructures comprising ZnO covered with Ga$_2$O$_3$ will significantly enhance the performance efficiency of UV photodetectors.

## Acknowledgments

Indian Institute of Technology Bombay, India is acknowledged by one of the authors (SP) for providing the Institute Postdoctoral fellowship. The authors are extremely grateful to Prof. M. S.

Ramachandra Rao (IIT Madras) and Dr. Tejendra Dixit (IIIT Kancheepuram) for their assistance with the synthesis and for providing the experimental facilities.

**Data availability statement –**

The raw/processed data required to reproduce these findings cannot be shared at this time as the data also forms part of an ongoing study.

**Compliance with ethical standards:**

Conflict of interest: The authors declare that they do not have any conflict of interest.

**Table of Contents**

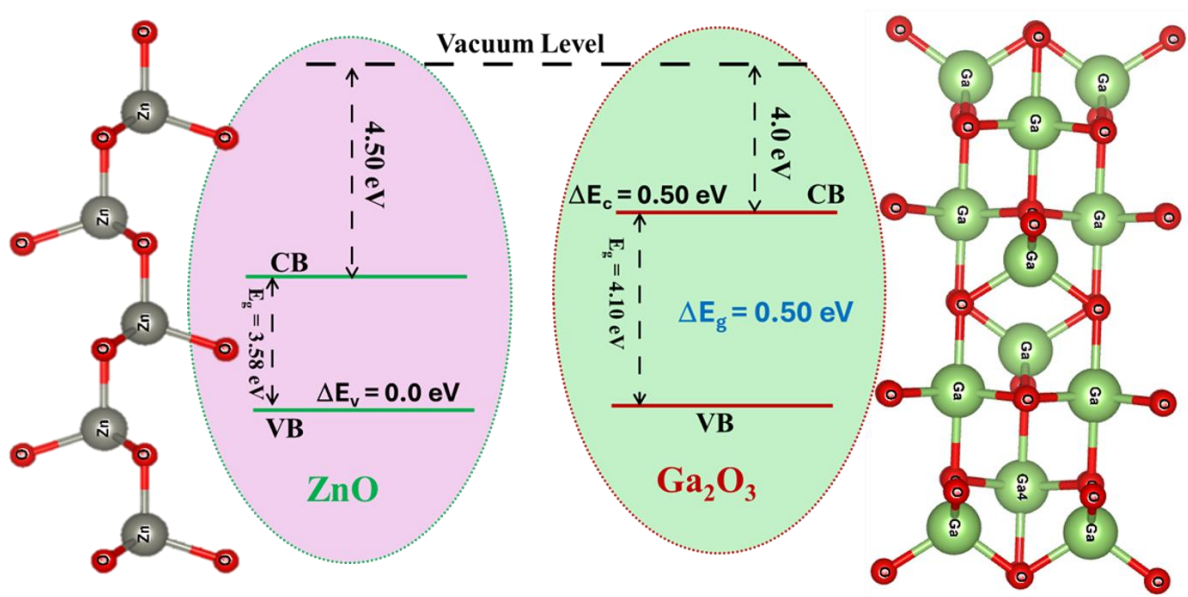

**References**

[1]  Pandey, S.; Shukla, A.; Tripathi, A. Highly sensitive and self powered ultraviolet photo detector based on ZnO nanorods coated with $TiO_2$, Sens. Actuators Phys. 350 (2023) 114112. https://doi.org/10.1016/j.sna.2022.114112.

[2]  Pandey, S.; Shukla, A.; Tripathi, A. Improvement in the performance and efficiency on self-deficient $CaTiO_3$: Towards sustainable and affordable new-generation solar cells, Mater. Today Commun. 35 (2023) 105575. https://doi.org/10.1016/j.mtcomm.2023.105575.

[3] Rajalekshmi, T.R.; Mishra, V.; Dixit, T.; Miryala, M.; Rao, M.S.R.; Sethupathi, K. Near white light and near-infrared luminescence in perovskite Ga:LaCrO$_3$, Scr. Mater. 210 (2022) 114449. https://doi.org/10.1016/j.scriptamat.2021.114449.

[4] Arige, S.; Mishra, V.; Miryala, M.; Rao, M.S.R.; Dixit, T. Plasmon-coupled sub-bandgap photoluminescence enhancement in ultra-wide bandgap CuO through hot-hole transfer, Opt. Mater. 134 (2022) 113149. https://doi.org/10.1016/j.optmat.2022.113149.

[5] Rai, H.M.; Saxena, S.K.; Mishra, V.; Warshi, M.K.; Kumar, R.; Sagdeo, P.R. Effect of Mn doping on dielectric response and optical band gap of LaGaO$_3$, Adv. Mater. Process. Technol. 3 (2017) 539–549. https://doi.org/10.1080/2374068X.2017.1349363.

[6] Singh, P.; Mishra, V.; Barman, S.; Balal, M.; Barman, S.R.; Singh, A.; Kumar, S.; Ramachandran, R.; Srivastava, P.; Ghosh, S. Role of H-bond along with oxygen and zinc vacancies in the enhancement of ferromagnetic behavior of ZnO films: An experimental and first principle-based study, J. Alloys Compd. 889 (2021) 161663. https://doi.org/10.1016/j.jallcom.2021.161663.

[7] Mishra, V.; Warshi, M.K.; Sati, A.; Kumar, A.; Mishra, V.; Kumar, R.; Sagdeo, P.R. Investigation of temperature-dependent optical properties of TiO$_2$ using diffuse reflectance spectroscopy, SN Appl. Sci. 1 (2019) 241. https://doi.org/10.1007/s42452-019-0253-6.

[8] Sumanth, A.; Mishra, V.; Pandey, P.; Rao, M.S.R.; Dixit, T. Investigations Into the Role of Native Defects on Photovoltaic and Spintronic Properties in Copper Oxide, IEEE Trans. Nanotechnol. 21 (2022) 522–527. https://doi.org/10.1109/TNANO.2022.3204587.

[9] Agrawal, J.; Dixit, T.; Palani, I.A.; Singh, V. Electron Depleted ZnO Nanowalls-Based Broadband Photodetector, IEEE Photonics Technol. Lett. 31 (2019) 1639–1642. https://doi.org/10.1109/LPT.2019.2940881.

[10] Maier-Flaig, F.; Rinck, J.; Stephan, M.; Bocksrocker, T.; Bruns, M.; Kübel, C.; Powell, A.K.; Ozin, G.A.; Lemmer, U. Multicolor Silicon Light-Emitting Diodes (SiLEDs), Nano Lett. 13 (2013) 475–480. https://doi.org/10.1021/nl3038689.

[11] You, D.; Xu, C.; Zhao, J.; Zhang, W.; Qin, F.; Chen, J.; Shi, Z. Vertically aligned ZnO/Ga$_2$O$_3$ core/shell nanowire arrays as self-driven superior sensitivity solar-blind photodetectors, J. Mater. Chem. C 7 (2019) 3056–3063. https://doi.org/10.1039/C9TC00134D.

[12] Raveesh, S.; Yadav, V.K.S.; Paily, R. CuO Single-Nanowire-Based White-Light Photodetector, IEEE Electron Device Lett. 42 (2021) 1021–1024. https://doi.org/10.1109/LED.2021.3081627.

[13] Jo, M.-S.; Song, H.-J.; Kim, B.-J.; Shin, Y.-K.; Kim, S.-H.; Tian, X.; Kim, S.-M.; Seo, M.-H.; Yoon, J.-B. Aligned CuO nanowire array for a high performance visible light photodetector, Sci. Rep. 12 (2022) 2284. https://doi.org/10.1038/s41598-022-06031-y.

[14] Abbasian, S.; Moshaii, A.; Vayghan, N.S.; Nikkhah, M. Ag Nanostructures Produced by Glancing Angle Deposition with Remarkable Refractive Index Sensitivity, Plasmonics 12 (2017) 631–640. https://doi.org/10.1007/s11468-016-0308-0.

[15] Agrawal, J.; Dixit, T.; Palani, I.A.; Singh, V. Highly Visible-Blind ZnO Photodetector by Customizing Nanostructures With Controlled Defects, IEEE Photonics Technol. Lett. 32 (2020) 1439–1442. https://doi.org/10.1109/LPT.2020.3031732.

[16] Chen, X.; Ahn, J.-H. Biodegradable and bioabsorbable sensors based on two-dimensional materials, J. Mater. Chem. B 8 (2020) 1082–1092. https://doi.org/10.1039/C9TB02519G.

[17] Bahri, M.; Yu, D.; Zhang, C.Y.; Chen, Z.; Yang, C.; Douadji, L.; Qin, P. Unleashing the potential of tungsten disulfide: Current trends in biosensing and nanomedicine applications, Heliyon 10 (2024) e24427. https://doi.org/10.1016/j.heliyon.2024.e24427.

[18] Alzahrani, H.; Sulaiman, K.; Muhammadsharif, F.F.; Abdullah, S.M.; Mahmoud, A.Y.; Bahabry, R.R.; Ab Sani, S.F. Effect of illumination intensity on a self-powered UV photodiode based on solution-processed NPD:Alq3 composite system, J. Mater. Sci. Mater. Electron. 32 (2021) 14801–14812. https://doi.org/10.1007/s10854-021-06034-x.

[19] Abdi, M.; Astinchap, B.; Khoeini, F. Electronic and thermodynamic properties of zigzag $MoS_2/MoSe_2$ and $MoS_2/WSe_2$ hybrid nanoribbons: Impacts of electric and exchange fields, Results Phys. 34 (2022) 105253. https://doi.org/10.1016/j.rinp.2022.105253.

[20] Chen, G.; Seo, J.; Yang, C.; Prasad, P. N. Nanochemistry and nanomaterials for photovoltaics, Chem. Soc. Rev. 42 (2013) 8304–8338. https://doi.org/10.1039/C3CS60054H.

[21] Combari, D.U.; Ramde, E.W.; Sourabie, I.; Zoungrana, M.; Zerbo, I.; Bathiebo, D.J. Performance Investigation of a Silicon Photovoltaic Module under the Influence of a Magnetic Field, Adv. Condens. Matter Phys. 2018 (2018) e6096901. https://doi.org/10.1155/2018/6096901.

[22] Grätzel, M. Perspectives for dye-sensitized nanocrystalline solar cells, Prog. Photovolt. Res. Appl. 8 (2000) 171–185. https://doi.org/10.1002/(SICI)1099-159X(200001/02)8:1<171::AID-PIP300>3.0.CO;2-U.

[23] Mishra, V.; Warshi, M.K.; Kumar, R.; Sagdeo, P.R.; Design and development of in-situ temperature dependent diffuse reflectance spectroscopy setup, J. Instrum. 13 (2018) T11003–T11003. https://doi.org/10.1088/1748-0221/13/11/T11003.

[24] Ferhati, H.; Djeffal, F.; Benhaya, A.-E.; Bendjerad, A. Giant Detectivity of ZnO-Based Self-Powered UV Photodetector by Inserting an Engineered Back Gold Layer Using RF Sputtering, IEEE Sens. J. 20 (2020) 3512–3519. https://doi.org/10.1109/JSEN.2019.2960271.

[25] Saha, R.; Karmakar, A.; Chattopadhyay, S. Enhanced self-powered ultraviolet photoresponse of ZnO nanowires/p-Si heterojunction by selective in-situ Ga doping, Opt. Mater. 105 (2020) 109928. https://doi.org/10.1016/j.optmat.2020.109928.

[26] Di Valentin, C.; Pacchioni, G.; Selloni, A. Reduced and n-Type Doped $TiO_2$: Nature of $Ti^{3+}$ Species, J. Phys. Chem. C 113 (2009) 20543–20552. https://doi.org/10.1021/jp9061797.

[27] von Wenckstern, H.; Kaidashev, E.M.; Lorenz, M.; Hochmuth, H.; Biehne, G.; Lenzner, J.; Gottschalch, V.; Pickenhain, R.; Grundmann, M. Lateral homogeneity of Schottky contacts on n-type ZnO, Appl. Phys. Lett. 84 (2004) 79–81. https://doi.org/10.1063/1.1638898.

[28] Xie, Y.; Nie, Y.; Zheng, Y.; Luo, Y.; Zhang, J.; Yi, Z.; Zheng, F.; Liu, L.; Chen, X.; Cai, P.; Wu, P. The influence of β-$Ga_2O_3$ film thickness on the optoelectronic properties of β-$Ga_2O_3$@ZnO nanocomposite heterogeneous materials, Mater. Today Commun. 29 (2021) 102873. https://doi.org/10.1016/j.mtcomm.2021.102873.

[29] Saha, R.; Chakrabarti, S.; Karmakar, A.; Chattopadhyay, S. Investigation of Yttrium (Y)-doped ZnO (Y:ZnO)–$Ga_2O_3$ core-shell nanowire/Si vertical heterojunctions for high-performance self-biased wideband photodetectors, J. Mater. Sci. Mater. Electron. 34 (2023) 759. https://doi.org/10.1007/s10854-023-10148-9.

[30] Pearton, S.J.; Yang, J.; Cary, P.H.; Ren, F.; Kim, J.; Tadjer, M.J.; Mastro, M.A. A review of $Ga_2O_3$ materials, processing, and devices, Appl. Phys. Rev. 5 (2018) 011301. https://doi.org/10.1063/1.5006941.

[31] Kong, W.-Y.; Wu, G.-A.; Wang, K.-Y.; Zhang, T.-F.; Zou, Y.-F.; Wang, D.-D. Luo, L.-B. Graphene-β-$Ga_2O_3$ Heterojunction for Highly Sensitive Deep UV Photodetector Application, Adv. Mater. 28 (2016) 10725–10731. https://doi.org/10.1002/adma.201604049.

[32] Zhang, J.; Dong, P.; Dang, K.; Zhang, Y.; Yan, Q.; Xiang, H.; Su, J.; Liu, Z.; Si, M.; Gao, J.; Kong, M.; Zhou, H.; Hao, Y. Ultra-wide bandgap semiconductor $Ga_2O_3$ power diodes, Nat. Commun. 13 (2022) 3900. https://doi.org/10.1038/s41467-022-31664-y.

[33] Aftab, S.; Abbas, A.; Iqbal, M.Z.; Hussain, S.; Kabir, F.; Akman, E.; Xu, F.; Hegazy, H.H. Recent advances in nanomaterials based biosensors, TrAC Trends Anal. Chem. 167 (2023) 117223. https://doi.org/10.1016/j.trac.2023.117223.

[34] Xie, K.; Umezawa, N.; Zhang, N.; Reunchan, P.; Zhang, Y.; Ye, J. Self-doped $SrTiO_{3-\delta}$ photocatalyst with enhanced activity for artificial photosynthesis under visible light, Energy Environ. Sci. 4 (2011) 4211–4219. https://doi.org/10.1039/C1EE01594J.

[35] Gottschalk, F.; Sonderer, T.; Scholz, R. W.; Nowack, B. Modeled Environmental Concentrations of Engineered Nanomaterials ($TiO_2$, ZnO, Ag, CNT, Fullerenes) for Different Regions, Environ. Sci. Technol. 2009, 43, 24, 9216–9222.

[36] Guo, S.; Kang, S.; Feng, S.; Lu, W. MXene-Enhanced Deep Ultraviolet Photovoltaic Performances of Crossed $Zn_2GeO_4$ Nanowires, J. Phys. Chem. C 124 (2020) 4764–4771. https://doi.org/10.1021/acs.jpcc.0c01032.

[37] Liu, Q.; Zhou, Y.; Kou, J.; Chen, X.; Tian, Z.; Gao, J.; Yan, S.; Zou, Z. High-yield synthesis of ultralong and ultrathin $Zn_2GeO_4$ nanoribbons toward improved photocatalytic reduction of $CO_2$ into renewable hydrocarbon fuel, J. Am. Chem. Soc. 132 (2010) 14385–14387. https://doi.org/10.1021/ja1068596.

[38] Therese, M.J.; Mishra, V.; Rao, M.S.R.; Dixit, T. First-Principles Calculations to Establish the Functionality of Self-Connected Point-Defect Migrations in n-ZnO- and p-CuO-Based Memristive Devices, IEEE Trans. Electron Devices 70 (2023) 6026–6033. https://doi.org/10.1109/TED.2023.3317000.

[39] Sumanth, A.; Mishra, V.; Rao, M.S. R.; Dixit, T. Interface Analysis of CuO/ZnO Heterojunction for Optoelectronic Applications: An Experimental and Simulation Study, Phys. Status Solidi A 220 (2023) 2300256. https://doi.org/10.1002/pssa.202300256.

[40] Göpel, W.; Schierbaum, K.D. SnO$_2$ sensors: current status and future prospects, Sens. Actuators B Chem. 26 (1995) 1–12. https://doi.org/10.1016/0925-4005(94)01546-T.

[41] Pandey, S.; Shukla, A.; Tripathi, A. Effect of pressure on electrical and optical properties of metal doped TiO$_2$, Opt. Mater. 133 (2022) 112875. https://doi.org/10.1016/j.optmat.2022.112875.

[42] Laks, D.B.; Van de Walle, C.G.; Neumark, G.F.; Blöchl, P.E.; Pantelides, S.T. Native defects and self-compensation in ZnSe, Phys. Rev. B 45 (1992) 10965–10978. https://doi.org/10.1103/PhysRevB.45.10965.

[43] Pejova, B.; Abay, B.; Bineva, I. Temperature Dependence of the Band-Gap Energy and Sub-Band-Gap Absorption Tails in Strongly Quantized ZnSe Nanocrystals Deposited as Thin Films, J. Phys. Chem. C 114 (2010) 15280–15291. https://doi.org/10.1021/jp102773z.

[44] Hu, Q.; Huang, J.; Li, G.; Chen, J.; Zhang, Z.; Deng, Z.; Jiang, Y.; Guo, W.; Cao, Y. Effective water splitting using CuOx/TiO$_2$ composite films: Role of Cu species and content in hydrogen generation, Appl. Surf. Sci. 369 (2016) 201–206. https://doi.org/10.1016/j.apsusc.2016.01.281.

[45] Pandey, S.; Shukla, A.; Tripathi, A. Elucidating the influence of native defects on electrical and optical properties in semiconducting oxides: An experimental and theoretical investigation, Comput. Mater. Sci. (2021) 111037. https://doi.org/10.1016/j.commatsci.2021.111037.

[46] Kumar, S.; Nandi, S.; Mishra, V.; Shukla, A.; Misra, A. Anomalous electrochemical capacitance in Mott-insulator titanium sesquioxide, J. Mater. Chem. A 10 (2022) 7314–7325. https://doi.org/10.1039/D1TA10262A.

[47] Singh, P.; Ghosh, S.; Mishra, V.; Barman, S.; Barman, S. R.; Singh, A.; Kumar, S.; Li, Z.; Kentsch, U.; Srivastava, P. Tuning of ferromagnetic behavior of GaN films by N ion implantation: An experimental and first principle-based study, J. Magn. Magn. Mater. 523 (2021) 167630. https://doi.org/10.1016/j.jmmm.2020.167630.

[48] Arige, S.; Mishra, V.; Miryala, M.; Rao, M.S.R.; Dixit, T. Plasmon-coupled sub-bandgap photoluminescence enhancement in ultra-wide bandgap CuO through hot-hole transfer, Opt. Mater. 134 (2022) 113149. https://doi.org/10.1016/j.optmat.2022.113149.

[49] Samat, M.H.; Taib, M.F.M.; Jaafar, N.K.; Hassan, O.H.; Yahya, M.Z.A.; Ali, A.M.M. First-principles studies on phase stability of $TiO_2$ by using GGA+U calculations, AIP Conf. Proc. 2030 (2018) 020058. https://doi.org/10.1063/1.5066699.

[50] Rajalekshmi, T.R.; Mishra, V.; Dixit, T.; Sagdeo, P.R.; Rao, M.S.R.; Sethupathi, K. Study of energy gaps and their temperature-dependent modulation in $LaCrO_3$: A theoretical and experimental approach, J. Appl. Phys. 133 (2023) 233104. https://doi.org/10.1063/5.0152325.

[51] Mishra, V.; Sati, A.; Warshi, M K.; Phatangare, A. B.; Dhole, S.; Bhoraskar, V. N.; Ghosh, H.; Sagdeo, A.; Mishra, V. Kumar R.; Sagdeo, P. R.; Effect of electron irradiation on the optical properties of $SrTiO_3$: An experimental and theoretical investigations, Mater. Res. Express 5 (2018) 036210. https://doi.org/10.1088/2053-1591/aab6f5

[52] Agrawal, J.; Dixit, T.; Palani, I. A.; Singh, V. Development of Reliable and High Responsivity ZnO-Based UV-C Photodetector, in IEEE Journal of Quantum Electronics, vol. 56, no. 1, pp. 1-5, Feb. 2020, Art no. 4000105, doi: 10.1109/JQE.2019.2957584.

[53] Dixit, T.; Agrawal, J.; Ganapathi, K. L.; Singh, V.; Rao, M. S. R. High-Performance Broadband Photo-Detection in Solution-Processed $ZnO$-$ZnCr_2O_4$ Nanowalls, in IEEE Electron Device Letters, vol. 40, no. 7, pp. 1143-1146, July 2019, doi: 10.1109/LED.2019.2916628.